\documentclass[twocolumn,showpacs,preprintnumbers,amsmath,amssymb]{revtex4}
\usepackage{graphicx}
\usepackage{dcolumn}
\usepackage{caption2}
\usepackage{bm}

\begin{document}

\preprint{APS/123-QED}

\title{Von Neumann entropy and localization-delocalization transition
 of electron states in quantum small-world networks}

\author {Longyan Gong $^{1,2}$}
\author{Peiqing Tong  $^{2}$}\thanks{Corresponding author.\\
 Email address:pqtong@pine.njnu.edu.cn, lygong@njupt.edu.cn}

\affiliation{
 $^{1}$ Department of Mathematics and
Physics, Nanjing University of Posts and Telecommunications,
Nanjing, Jiangsu 210003, P.R.China
 \\ $^{2}$ Department of Physics, Nanjing Normal University, Nanjing, Jiangsu 210097, P.R.China}

\date{today}
\begin{abstract}
The von Neumann entropy for an electron in periodic, disorder and
quasiperiodic quantum small-world networks(QSWNs) are studied
numerically. For the disorder QSWNs, the derivative of the spectrum
averaged von Neumann entropy is maximal at a certain density of
shortcut links $p^*$, which can be as a signature of the
localization-delocalization transition of electron states. The
transition point $p^*$ is agreement with that obtained by the level
statistics method. For the quasiperiodic QSWNs, it is found that
there are two regions of the potential parameter. The behaviors of
electron states in different regions are similar to that of periodic
and disorder QSWNs, respectively.
\end{abstract}
\maketitle

\section{Introduction}

Recently the small-world networks (SWNs) \cite{wa98}have attracted
much attention since it can mimic social and biological networks,
Internet connections, airline flights and other complex networks.
Well-established classical models have been numerically and
analytically investigated, which focused on the crossover behavior
\cite{ba99}, the scaling properties \cite{ku00,ne99} and the
percolation of the dynamic processes \cite{ne99} in the model,
etc. Very recently Zhu and Xiong have generalized the SWNs to a
quantum version by regarding the bonds as quantum hopping links
for the motion of an electron and investigated the
localization-delocalization transition of electron
states\cite{zh00}. Until now the transition point is found only by
using the level statistics method combined with the finite-size
scaling method\cite{zh00,ch01,sa05}. However the finite-size
scaling method is not suitable for SWNs with high connections
since the number of connections grows very rapidly with the SWNs
size\cite{sa05}. The level statistics method is successful in the
location of the metal-insulator transition in disorder systems
\cite{sh93,ev95}, but it is not suitable in quasiperiodic systems
because the level spacing distribution \cite{ma86} can not be
always written as the crossover of Poisson distribution and
Wigner-Dyson distribution.

On the other hand, the connection between the entanglement (such
as von Neumann entropy) and localization properties of eigenstates
is revealed recently. By measuring the von Neumann entropy, the
local entanglement was studied at the ground state in the Hubbard
model for the dimer case \cite{za02} and in the extended Hubbard
model for different band filling \cite{gu04}. It is found that the
von Neumann entropy is suitable to describe quantum phase
transition \cite{gu04} and analyze the interplay between itinerant
and localized feature\cite{za02}.

In this paper, we study von Neumann entropy for an electron moving
in periodic, random and  quasiperiodic quantum small-world
networks(QSWNs), respectively. In periodic QSWNs, there are no
localization-delocalization transitions because all eigenstates
are extended. In random QSWNs, we find the spectrum averaged von
Neumann entropy is a suitable quantity to analyze the
localization-delocalization transition of electron states.
Finally, we propose quasiperiodic QSWNs based on one-dimensional
Harper model. With the help of von Neumann entropy, we find that
there are two regions of the potential parameter in the model. The
behaviors of electron states in different regions are similar to
that of periodic and disorder QSWNs, respectively.

The paper is organized as follows. In the next section we describe
the QSWNs model and the measure of entanglement. In Sec.III we
present numerical results for different QSWNs. Finally, in Sec.
IV, the conclusions are given.

\section{model and von Neumann entropy}

\begin{figure}
\includegraphics[width=2.5in]{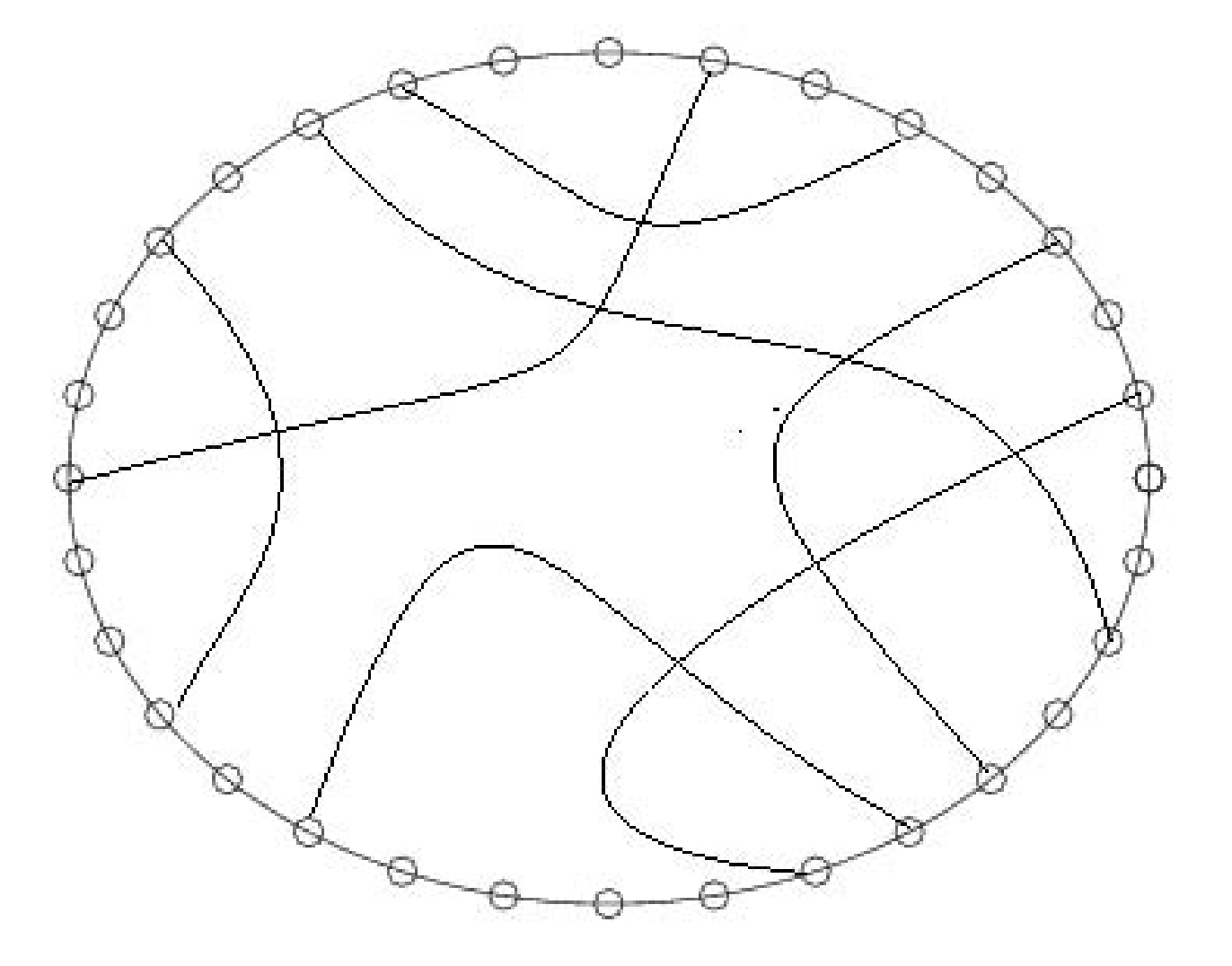}
\caption{An example of a small-world graph with vertices $N=32$
and shortcut links $pN=7$ .}\label{fig1}
\end{figure}  

We consider a circular graph with $N$ vertices. Each vertex is
linked with (direct connections) its two nearest-neighbors. To
this graph, $pN$ shortcut links (connecting $2pN$ vertices) are
additionally added between random pairs of vertices without direct
connections (Fig.\ref{fig1})\cite{ne99}. Here $p$ is the density
of shortcuts links.

The tight-binding Hamiltonian of an electron in QSWNs can be
written as
\begin{eqnarray}
H =H_0+H_1, \label{form1}
\end{eqnarray} 
where
\begin{eqnarray}
H_0 =t\sum\limits_{n = 1}^N {(c_n^ \dag  c_{n + 1}  + c_{n + 1}^
\dag c_n)} + \sum\limits_{n = 1}^N {\varepsilon_n c_n^ \dag c_n},
\label{form2}
\end{eqnarray} 
and \begin{eqnarray}
H_1=t_1\sum\limits_{k=1}^{pN}\sum\limits_{n=1}^N
\sum\limits_{m=1}^N{(c_n^ \dag  c_m  + c_m^ \dag
c_n)}\delta_{n,n_k}\delta_{m,m_k}.\label{form3}
\end{eqnarray}
Here $t$ is a nearest-neighbor hopping integral, $c_n^\dag$
($c_n$) the creation(annihilation) operator of \emph{n}th site,
$\varepsilon_n$ the on-site potential. It is clear that the $H_0$
defines a one-dimensional tight-binding model.  $t_1$ is hopping
integral for shortcut links; $\{n_k,m_k\}$( here we restrict that
$n_k<m_k$) are the pairs of vertices connected by a shortcut link
and  the number of all pairs is $pN$. Theoretically
$(pN)_{max}=N(N-3)/2$. Here we only study small value $p$.

Let $\left|n \right\rangle \equiv c_n^ \dag\left| 0 \right\rangle$,
the general eigenstate of an electron is
\begin{equation}
\left| \alpha  \right\rangle  = \sum\limits_{n = 1}^N {\psi^\alpha
_n } \left| n \right\rangle  = \sum\limits_{n = 1}^N {\psi^\alpha
_n c_n^ \dag } \left| 0 \right\rangle,\label{form5}
\end{equation} 
where ${\psi^\alpha _n }$ is the amplitude of wave function
$\alpha$ at the \emph{n}th site.

The general definition of entanglement is based on the von Neumann
entropy \cite{be96}. For an electron in the system, there are two
possible local states at each site, $\left| 1 \right\rangle_n$ and
$\left| 0 \right\rangle_n$, corresponding to the state with(out)
an electron at the \emph{n}th site, respectively.
The local density matrix $\rho_n$ is
defined \cite{za02,gu04} by
\begin{equation}
\rho_n= z_n\left| {1} \right\rangle{_n}{_n}\left\langle {1}
\right| + (1-z_n)\left| {0} \right\rangle{_n}{_n}\left\langle {0}
\right|,\label{form9}
\end{equation} 
where $z_n=\left\langle \alpha  \right|c_n^ \dag  c_n \left|
\alpha \right\rangle=\left|\psi^\alpha _n \right|^2$ is the local
occupation number at the \emph{n}th site. The corresponding von
Neumann entropy is
\begin{equation}
E^\alpha_{vn}=-z_n\log_2z_n-(1-z_n)\log_2(1-z_n),\label{form10}
\end{equation}
which measures the entanglement of states on the \emph{n}th site
with that on the remaining $N-1$ sites. It is called the local
entanglement for it exhibits the correlations between a site and
all the other sites of the system \cite{za02,gu04}. Generally the
$E^\alpha_{vn}$ is a function of $n$. We define the von Neumann
entropy of system at $\alpha$ eigenstate is
\begin{equation}
E^\alpha_v= \frac{1}{N} \sum\limits_{n=1}^N
{E^\alpha_{vn}}.\label{form11}
\end{equation} 
The definition (\ref{form11}) shows that for an extended state
that $\psi^\alpha_{n}=\frac{1}{\sqrt{N}}$ for all $n$,
 $E^\alpha_v=-\frac{1}{N}\log_2 \frac{1}{N}- (1-\frac{1}{N})\log_2
(1-\frac{1}{N}) \approx \frac{1}{N}\log_2{N}$ at $N
\longrightarrow\infty$, and for a localized state that
$\psi^\alpha_n=\delta_{nn^\circ}$( $n^ \circ$ is a given site ) ,
$E^\alpha_v=0$. In the paper all the values of $E^\alpha_v$ and
$E^\alpha_{vn}$ are scaled by $\frac{1}{N}\log_2{N}$. From the two
examples, we know the scaled $E^\alpha_v$ is near $1$ when
eigenstates are extended, and near zero when eigenstates are
localized. Henceforth, we omit ``scaled'' for simplicity.

As a further gross measure we also average over all the
eigenstates, i.e., the spectrum averaged von Neumann entropy
\begin{equation}
\langle E_v \rangle = \frac{1}{M}\sum\limits_{\alpha} {E^\alpha_v
},\label{form12}
\end{equation} 
where $M$ is the number of all eigenstates.\\

\begin{figure}[!ht]
\includegraphics[width=2.5in]{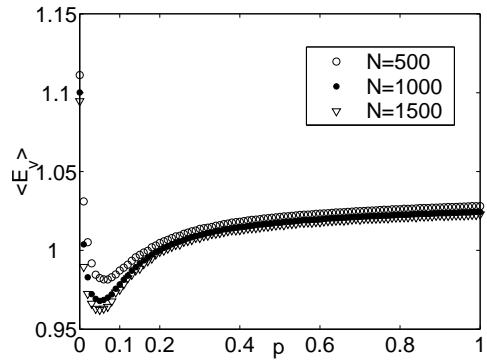} \caption{The spectrum averaged
von Neumann entropy $\langle E_v \rangle$  varying with  $p$ at
different sizes $N$ for periodic QSWNs.}\label{fig2}
\end{figure}

\section{numerical results}

In our numerical calculations, the Hamiltonian is obtained
according to the formulas (\ref{form1})---(\ref{form3}) for finite
systems. The shortcut terms are generated randomly based on the
formula(\ref{form3}). We directly diagonalize the Hamiltonian and
obtain $N$ eigenvalues $E_\alpha$ and corresponding eigenvectors
$\left| \alpha \right\rangle$. From the formulas
(\ref{form9})---(\ref{form12}), we obtain the spectrum averaged
von Neumann entropy $\langle E_v \rangle$ for one realization of
QSWNs. The results are averaged over many realizations of QSWNs.

\subsection{Periodic QSWNs}

For periodic QSWNs, the on-site potential $\varepsilon_n$ is
assumed to be uniform and set equal to zero. Without loss
generality and for simplicity, we set $t=t_1=1$ in all our
numerical calculations. The Fig.\ref{fig2} shows the spectrum
averaged von Neumann entropy $\langle E_v \rangle$ changing with
$p$ at $N=500, 1000$ and $1500$ , respectively. Averages are done
for 300 ,200, and 100 random configurations (positions of shortcut
links ) at $N=500, 1000$ and $1500$, respectively.  The results
are similar for more random configurations. From the figure, we
can see that $\langle E_v\rangle$ is close to $1$ for all $p$,
which means that  all states are extended and there is no
localization-delocalization transition in the systems. For $p=0$,
i.e. in the absence of shortcuts, the model is a one-dimensional
periodic potential system. The energy eigenstates are always
extended  due to the Bloch theorem. The random shortcut terms can
cause the long-range hopping and off-diagonal disorder effects.
The long-range hopping tends to delocalize the states, therefore
the extensive properties of the eigenstates are not changed by the
presence of random shortcut terms.
We also find there is small decreases of $\langle E_v\rangle$ for
{\sl very small $p$}($p<0.05$), which is due to localization
effects of the off-diagonal disorder caused by the random shortcut
terms in the Hamiltonian.

\begin{figure}
\includegraphics[width=2.5in]{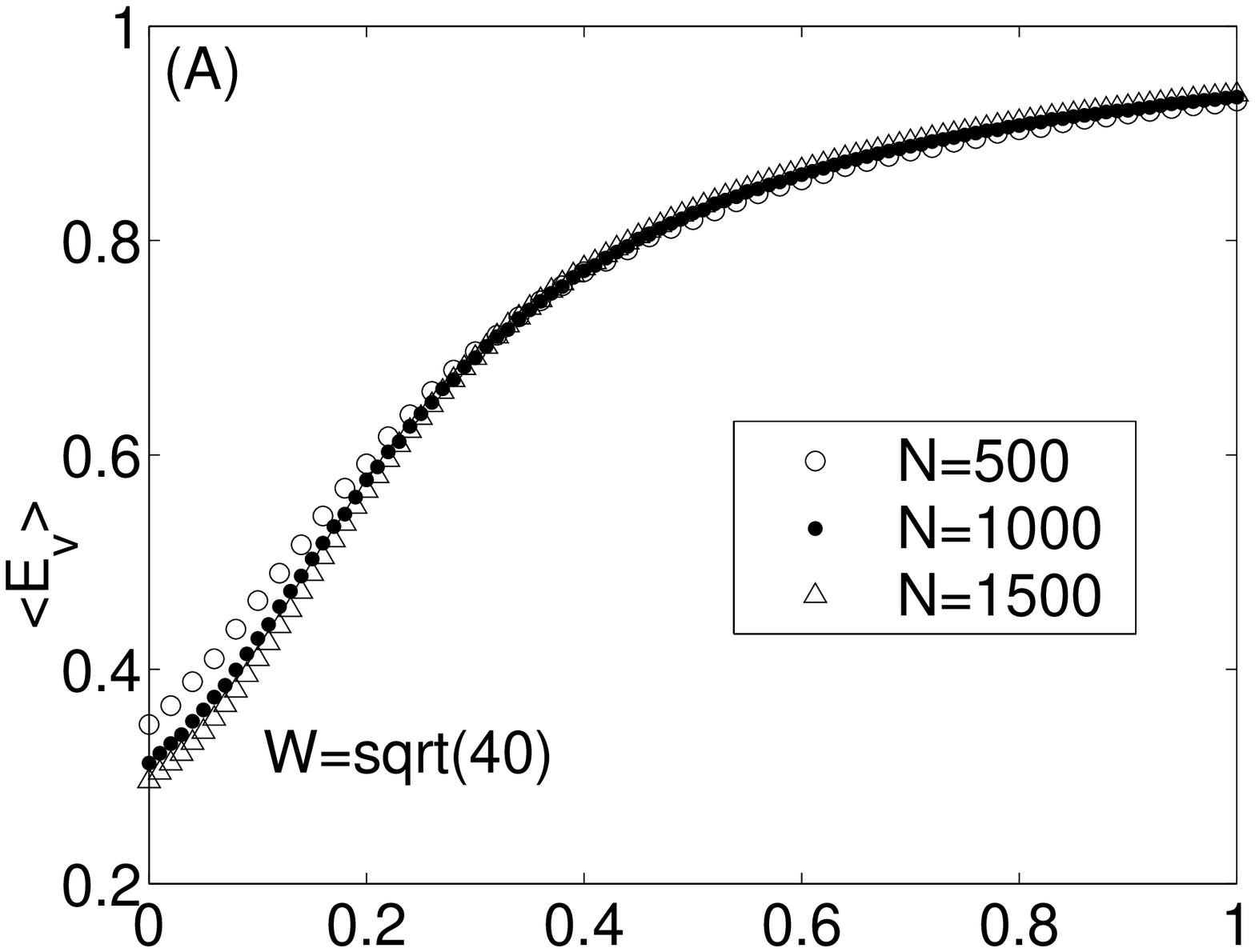}
\includegraphics[width=2.5in]{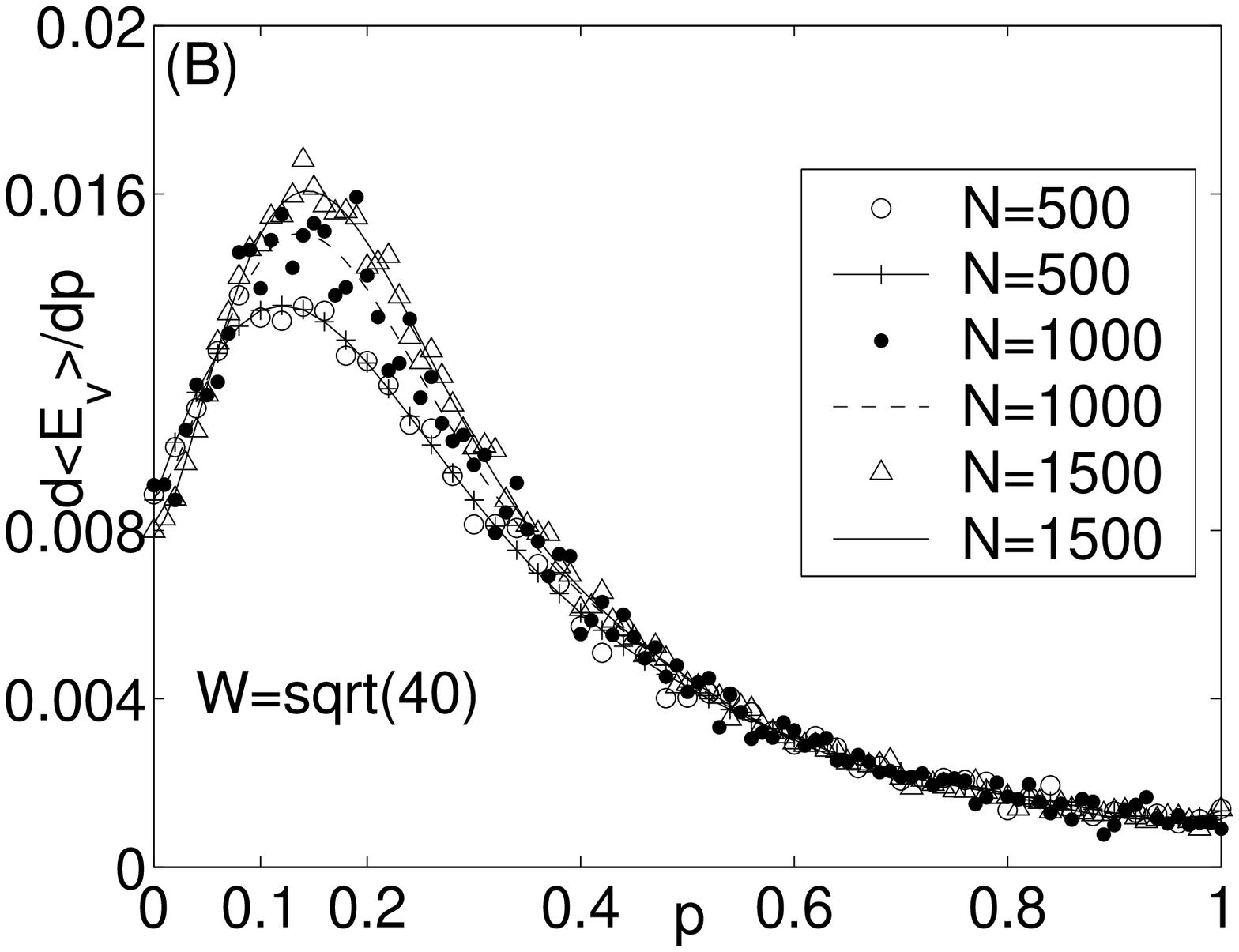}
\includegraphics[width=2.5in]{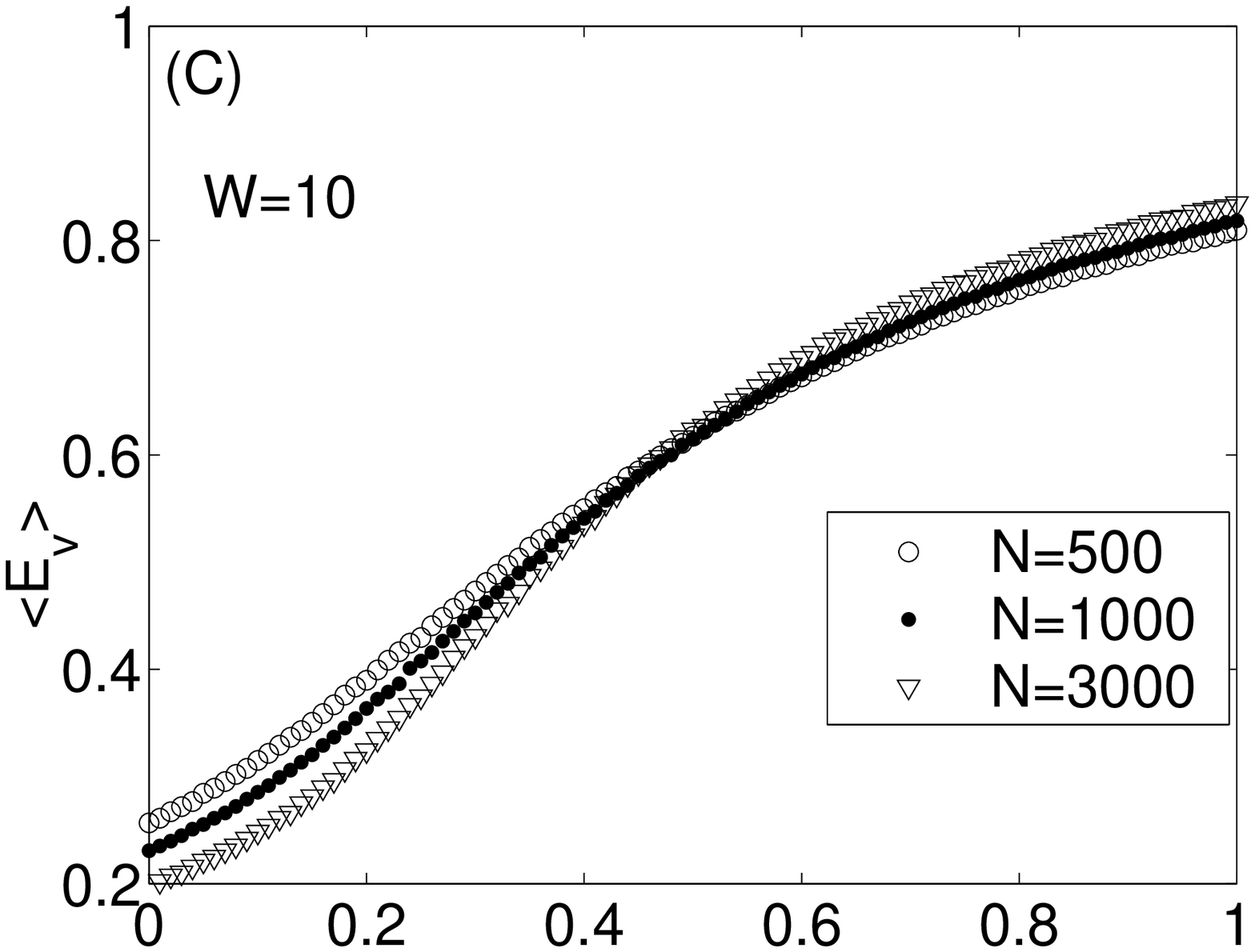}
\includegraphics[width=2.5in]{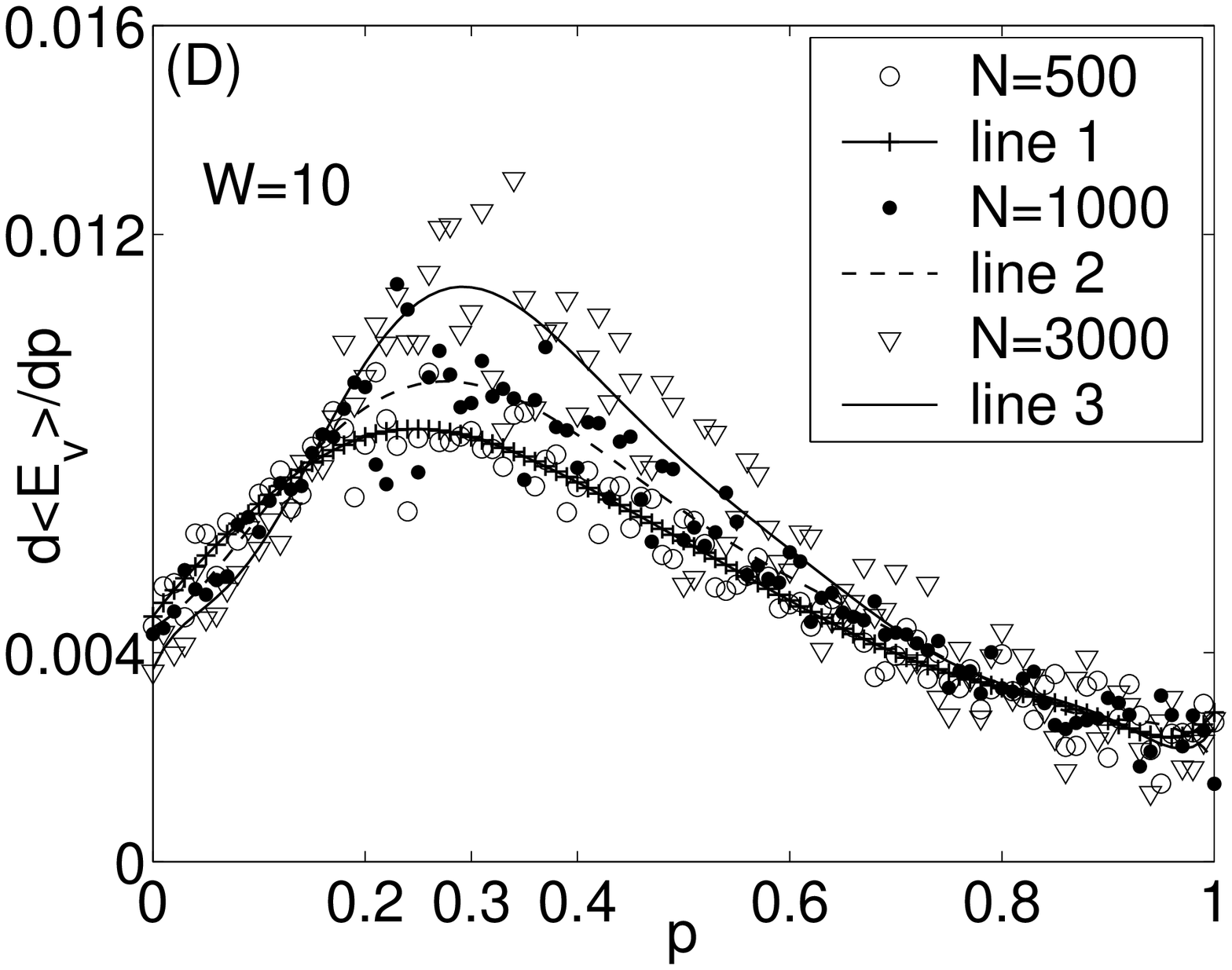} \caption{The spectrum averaged
von Neumann entropy $\langle E_v \rangle$ (A,C) and the derivative
$d\langle E_v \rangle/dp$ (B,D) varying with  $p$ at different
sizes $N$ for $W=\sqrt{40}$ (A,B) and $W=10$ (C,D), respectively.
Lines in figures are polynomial fitted for corresponding data. The
number of disorder realizations (positions of shortcut links and
on-site disorder potential)is $300$, $200$, $100$ and $20$ for
$N=500$, $1000$, $1500$ and $3000$, respectively. }\label{fig3}
\end{figure}  

\subsection{Disordered QSWNs}
For disordered QSWNs, the on-site potential $\varepsilon_n$ are
random variables homogeneously distributed with $[-W/2;W/2]$. Here
the $W$ characterizes the degree of on-site disorder as in the
Anderson model \cite{an58}.  By using the level statistics method,
it has been found that a transition from Possion
statistics(localized phase) to Wigner-Dyson statistics
(delocalized phase) takes place at $ p_c\approx
\frac{1}{{400}}(W/t)^2 $ for weak disorder, i.e., $W/t$ is small
\cite{ch01}.

In Fig.\ref{fig3} we show the spectrum averaged von Neumann
entropy $\langle E_v \rangle$ and the derivative $d\langle E_v
\rangle/dp$  varying with $p$ for $W=\sqrt{40}$ and $W=10$ at
different $N$, respectively.  From \ref{fig3} (A) and (C), it's
clear that $\langle E_v \rangle$ monotonically increases as $p$
becomes larger.  When $p=0$, the model is a one-dimensional
Anderson model \cite{an58}. For the model all states are
localized, so $\langle E_v \rangle$ are small($\langle E_v \rangle
\approx 0.3$ and $0.2$ at $W=\sqrt{40}$ and $10$, respectively).
When $p$ is large, delocalized states will be present due to the
long-range hopping, and $\langle E_v \rangle$ becomes large. From
Fig.\ref{fig3} (B) and (D), it is found that the derivative
$d\langle E_v \rangle/dp$ is maximal at $p^*\approx 0.1\sim0.15$
and $0.25\sim0.3$ at $W=\sqrt{40}$ and $W=10$, respectively. The
$p^*$ is agreement with the localization- delocalization
transition point $p_c$ obtained by the level statistics method
($p_c\approx 0.1$ at $W=\sqrt{40}$ and $p_c\approx 0.25$ at
$W=10$)\cite{ch01}. It is clear that the transition from localized
phase to delocalized phase can also be reflected from $d\langle
E_v \rangle/dp$. Therefore the von Neumann entropy is a suitable
quantity to analysis localized properties of electron states for
QSWNs.

\subsection{Quasiperiodic QSWNs }

\begin{figure}[!ht]
\includegraphics[width=2.5in]{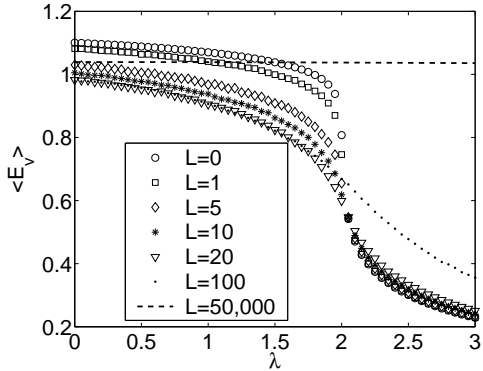}
 \caption{The spectrum
averaged von Neumann entropy $\langle E_v \rangle$ varying with
$\lambda$ at different shortcut links number $L$ . Here N=987 and
the number of random configurations (positions of shortcut links)
is $200$.}\label{fig4}
\end{figure}

\begin{figure}[!ht]
\includegraphics[width=2.5in]{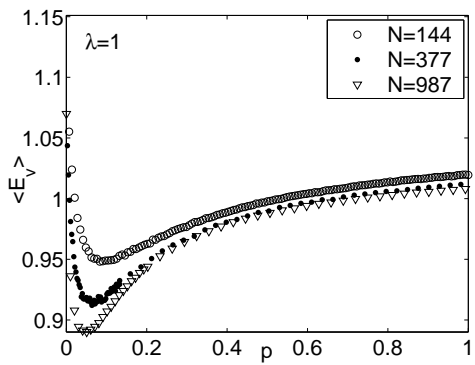}\caption{The spectrum averaged
von Neumann entropy $\langle E_v \rangle$ as a function of $p$ at
$\lambda=1$ for different system sizes. The number of random
configurations (positions of shortcut links) is $500$, $300$ and
$200$ for $N=144$, $377$ and $987$, respectively.}\label{fig5}
\end{figure} 

After the experimental discovery of the quasicrystals \cite{sh84}
and one-dimensional quasiperiodic superlattices \cite{me85}, many
experimental and theoretical works have been carried out on the
physical properties of quasiperiodic systems
\cite{so85,au80,ko83,os83,ko831,os84,th83}. Although these systems
lack translational invariance, they are perfect ordered. In this
sense, such systems can be regarded as intermediate between periodic
and random systems. The one of the most popular quasiperiodic
systems is Harper model. In the following, we propose a
quasiperiodic QSWN based on the Harper model and study the
properties of the eigenstates of an electron in this system.

For the Harper quasiperiodic QSWNs, we choose $\varepsilon_n=\lambda
cos(2\pi\sigma n)$ and $\sigma$ is irrational. The potential is
incommensurate with the underlying vertices. At $p = 0$ the model is
in fact the one dimensional well studied Harper model\cite{so85}.
Intensively analytical and numerical studies
\cite{so85,au80,ko83,os83,ko831,os84,th83} for the Harper model show
that for $\lambda <2$ the spectrum becomes absolute continues and
all  eigenstates are extended. For $\lambda
>2$ the spectrum is pure point and all eigenstates are
exponentially localized.  For $\lambda =2$ the situation gives the
metal-insulator transition at which the eigenstates are neither
extended nor localized but critical with a singular-continuous
multifractal spectrum.

As a typical case, we take $\sigma=(\sqrt{5}-1)/2$. In fact as is
customary in the context of quasiperiodic system, the value of
$\sigma$ may be approximated by the ratio of successive Fibonacci
numbers---$F_n=F_{n-2}+F_{n-1}$ . In this way, choosing
$\sigma=F_{n-1}/F_n\approx(\sqrt{5}-1)/2$ and system size $N=F_n$,
we can obtain the periodic approximant for the quasiperiodic
potential\cite{th83}

In Fig.\ref{fig4}, we plot the spectrum averaged von Neumann
entropy $\langle E_v \rangle$ varying with $\lambda$ for different
shortcut links number $L$( here $L=pN$). For $L=0$, $\langle E_v
\rangle$ is large at $\lambda<2$, while small at $\lambda>2$.
There is a sharp decrease in $\langle E_v \rangle$ for
$\lambda=2$, i.e., the absolute value of $d\langle E_v \rangle/dp$
is maximal at $\lambda=2$, so the metal-insulator transition can
be reflected from $\langle E_v \rangle$. When $L$ is small
($L\leq20$), those varyings properties of $\langle E_v \rangle$
are similar to that for $L=0$, which means that at small $L$, the
quasiperiodic on-site potentials rather than shortcut links play
an important role. When $L$ is large ($L$=100), the decrease in
$\langle E_v \rangle$ at $\lambda=2$ is not so sharp as that for
$L\leq20$. When $L$ is large enough (for example $L=50,000$),
$\langle E_v \rangle$ is almost same for all $\lambda$. At the
situation the SWNs is almost completely a random graph and the
on-site potential is not important.

For $\lambda<2$, the varying properties of $\langle E_v \rangle$
with $p$ are similar to that of periodic QSWNs. In Fig.\ref{fig5}
$\lambda=1$ is given as an example. It shows that on the whole,
for all $p$, $\langle E_v \rangle$ is near $1$, which means all
states are extended.

\begin{figure}
A)\includegraphics[width=2.5in]{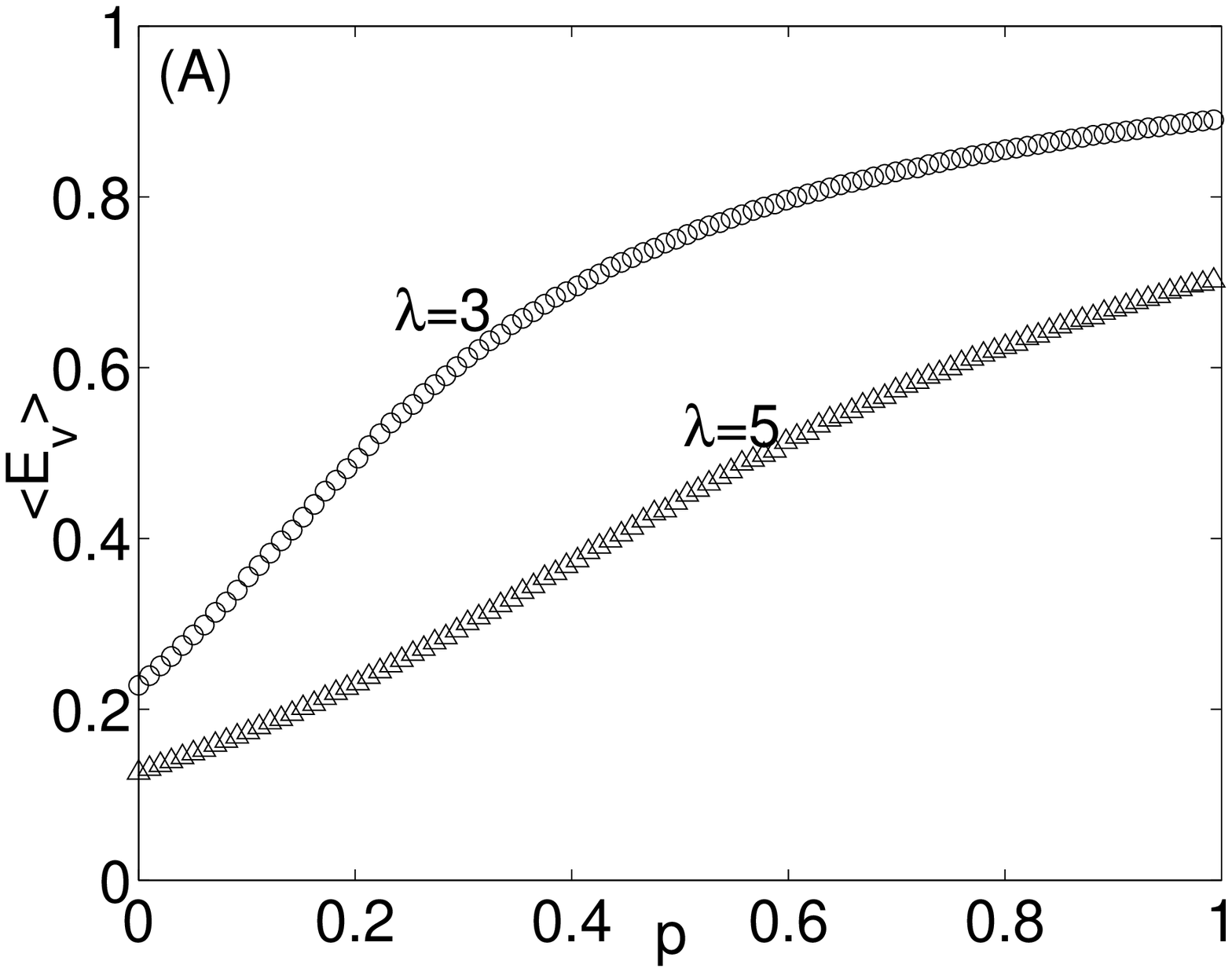}
(B)\includegraphics[width=2.5in]{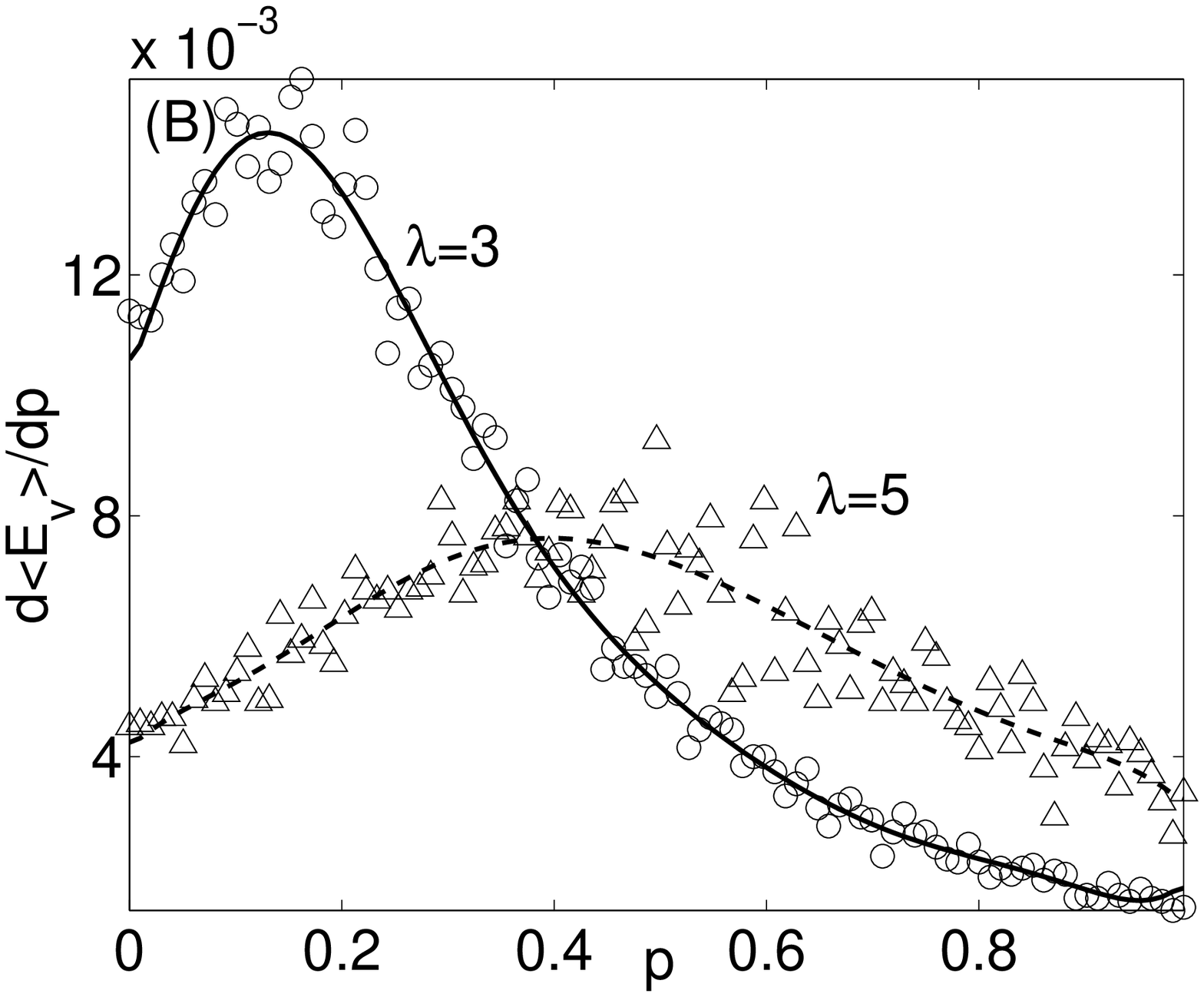} \caption{The spectrum
averaged von Neumann entropy $\langle E_v \rangle$ and $d\langle
E_v \rangle/dp$ varying with the density of shortcut links $p$ for
(A) and (B), respectively. Lines in figure is polynomial fitted
for corresponding data. Here $N=987$ and the number of random
configurations (positions of shortcut links) is $200$.
}\label{fig6}
\end{figure}

For $\lambda>2$, the varying properties of $\langle E_v \rangle$
and the derivative $d\langle E_v \rangle/dp$ with $p$ are similar
to those for disorder QSWNs. In Fig.\ref{fig6} $\lambda=3$ and $5$
are as examples.  The spectrum averaged von Neumann entropy
$\langle E_v \rangle$ and the derivative $d\langle E_v \rangle/dp$
with different $p$ are shown in Fig. \ref{fig6}(A) and (B),
respectively. It shows that $\langle E_v \rangle$ monotonically
increases as $p$ increases. The derivative $d\langle E_v
\rangle/dp$ is maximal at $p^*\approx 0.12$ and $0.4$ at
$\lambda=3$ and $5$, respectively, so the
localization-delocalization transition happens at $p^*$. This also
can be certified by the level statistics method.

\begin{figure}
(A)\includegraphics[width=2.5in]{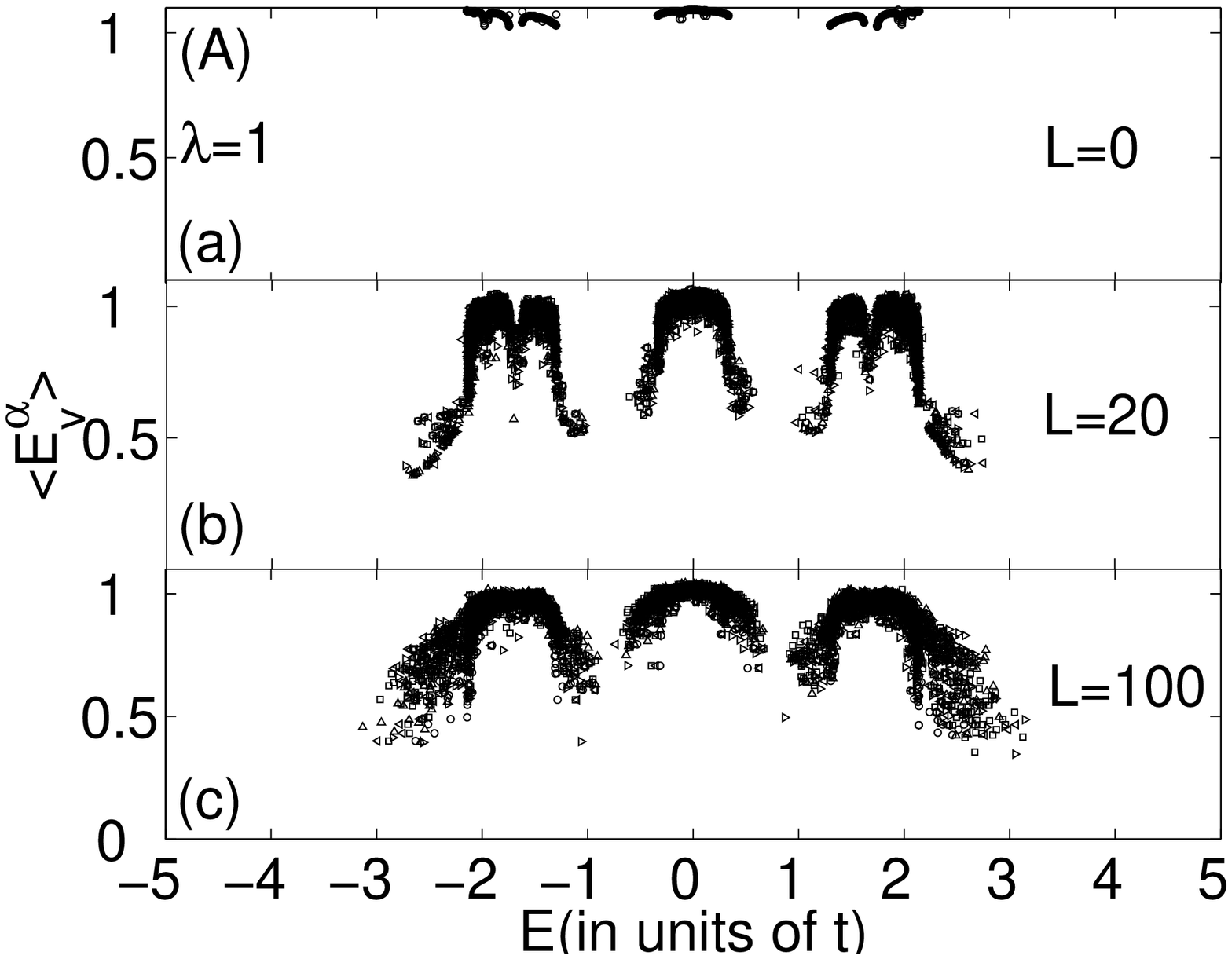}
(B)\includegraphics[width=2.5in]{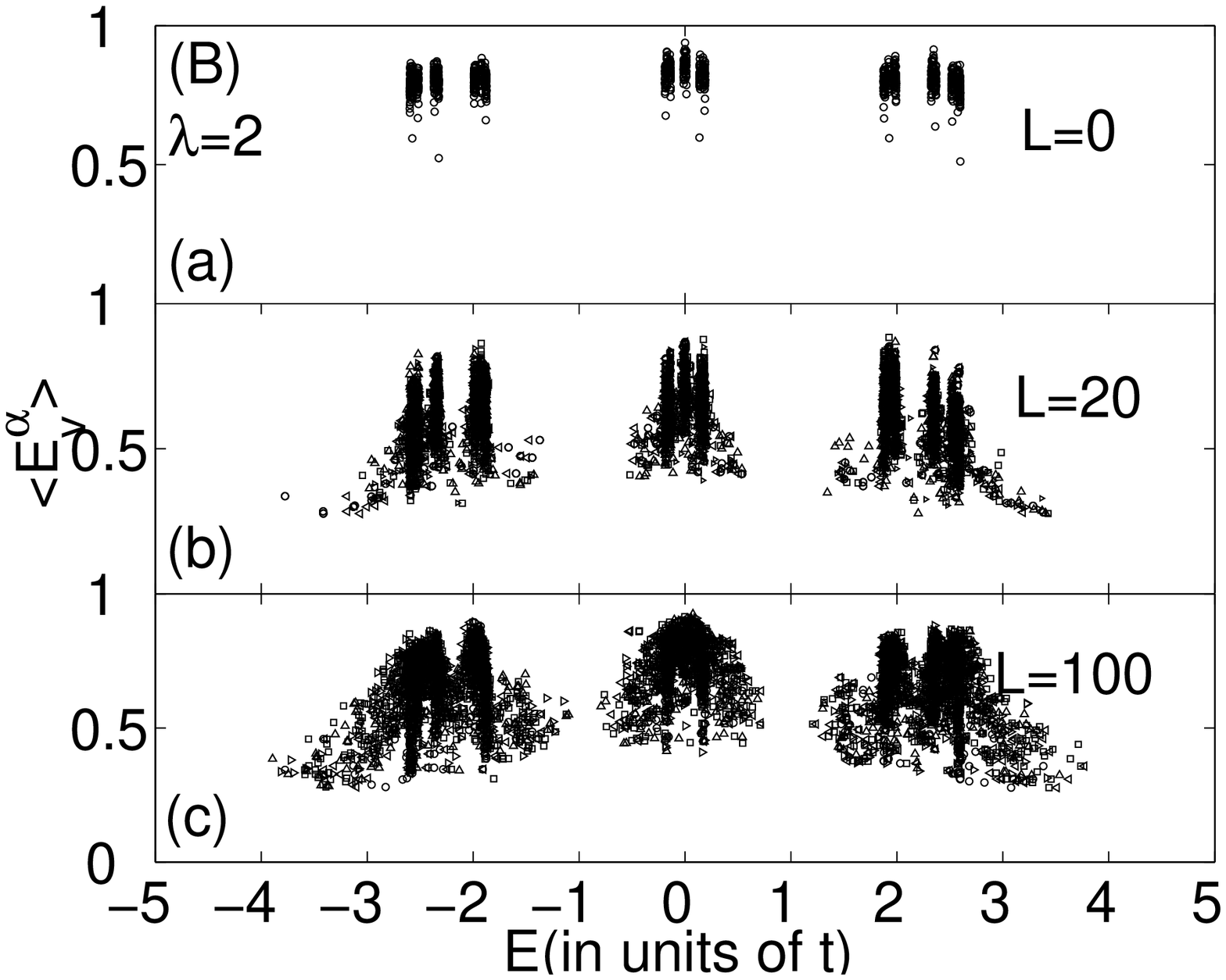}
(C)\includegraphics[width=2.5in]{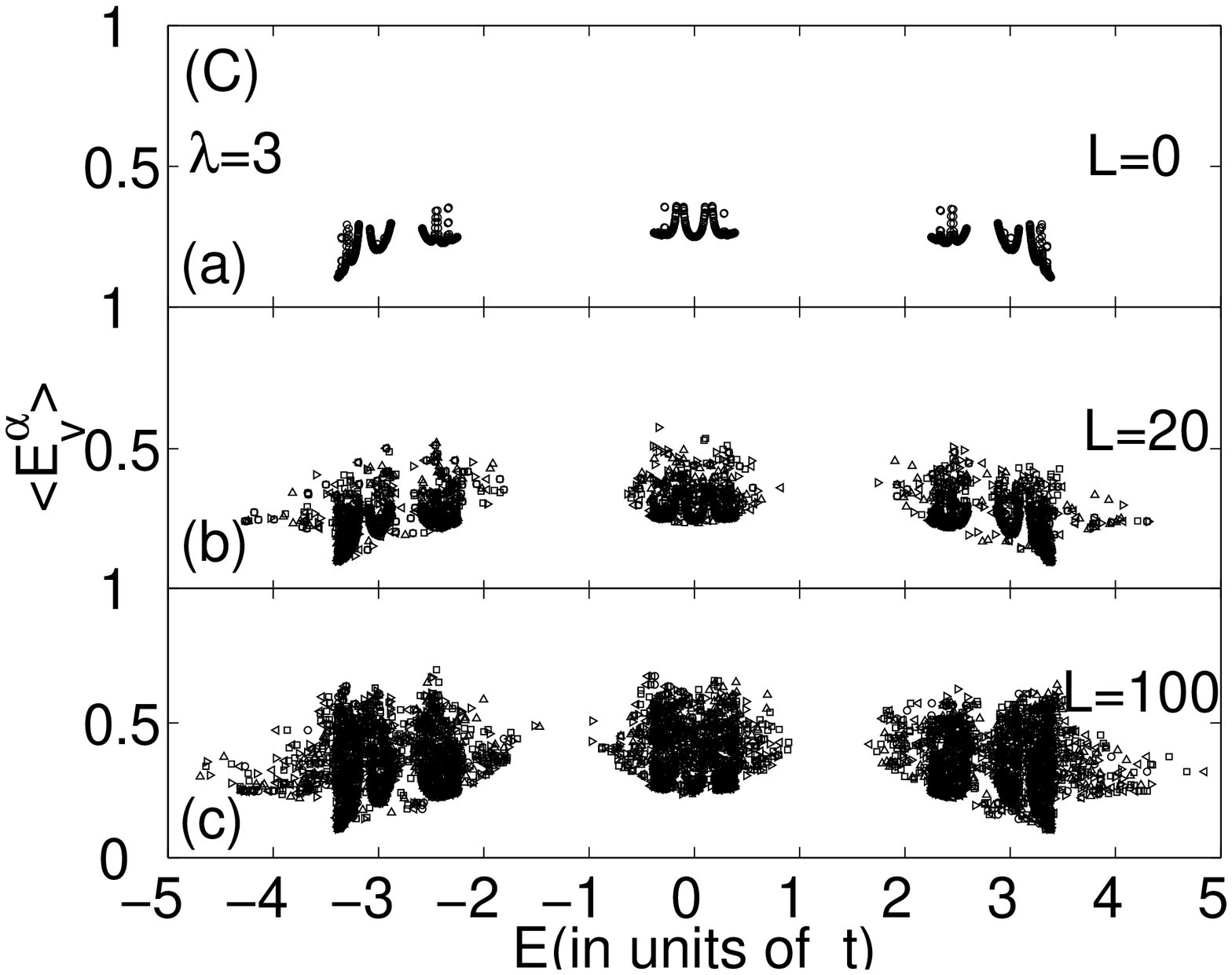} \caption{ Average von
Neumann entropy $\langle E_v^\alpha \rangle$ of the individual
eigenstate as functions of eigenenergies at $L=0,20$ and $100$ for
(A)$\lambda=1$, (B)$\lambda=2$ and (C)$\lambda=3$, respectively.
For $L>0$, the $\langle E_v^\alpha \rangle$ value for six random
configuration( positions of shortcut links ) of quasiperiodic
QSWNs are plotted together. Here $N=987$.}\label{fig7}
\end{figure}  

To understand the effect of shortcut links clearly, in
Fig.\ref{fig7}(A), (B) and (C) we plot the average von Neumann
entropy $\langle E_v^\alpha \rangle$ of the individual eigenstates
at $\lambda=1,2$ and $3$ for $L=0,20$ and $100$, respectively.
When $\lambda=1$, at $L=0$ all $\langle E_v^\alpha \rangle$ are
large (near $1$), which corresponds to that all eigenstates are
extended. At $L=20$, the subband created by the shortcut links
lies below the band bottom , above the band top and at the band
gap of that for $L=0$. In those new created subbands, $\langle
E_v^\alpha \rangle$ are obviously small comparing to that for
$L=0$, which means shortcut links can produce localized states at
the case. As $L$ increases to $100$ and the long-hopping becomes
more and more important, on the whole $\langle E_v^\alpha \rangle$
in the new created subbands are larger than that for $L=20$. When
$\lambda=3$, at $L=0$ all $\langle E_v^\alpha \rangle$ are small
comparing to that for $\lambda=1$ at $L=0$, which corresponds that
all eigenstates are localized. At $L>0$, $\langle E_v^\alpha
\rangle$ for most eigenstates  are large comparing with that at
$L=0$. At the situation the long-hopping due to shortcut links is
important and make many states more extended than that at $L=0$.
When $\lambda=2$, at $L=0$ the eigenstates are critical with a
singular-continuous multifractal spectrum. At the situation some
eigenstates have large $\langle E_v^\alpha \rangle$ and some have
small $\langle E_v^\alpha \rangle$. At $L=20$, $\langle E_v^\alpha
\rangle$ become larger at some eigenstates  and smaller at some
eigenstates due to the shortcut links. The spectrum averaged von
Neumann entropy $\langle E_v\rangle$ changes little. At $L=100$,
the long-hopping becomes important and leads many eigenstates have
large $\langle E_v^\alpha \rangle$ comparing to that for $L=20$.

\section{conclusions}

In detail, we study von Neumann entropy in periodic and disorder
QSWNs and find it is a suitable quantity to reflect
localization-delocalization transition of electron states. Then we
propose a quasiperiodic QSWN based on one-dimensional Harper model
and investigate it intensively by the measure of von Neumann
entropy. In the model, the quasiperiodic on-site potential, the
long-range hopping and off-diagonal disorder due to random
shortcut links determine the localization properties of electron
states. Those lead that the influence of shortcut links on von
Neumann entropy are different at two $\lambda$ regions($\lambda<2$
and $\lambda>2$ ). We found that when $\lambda<2$, we find that on
the whole, for all $p$, $\langle E_v \rangle$ is near $1$, which
means all states are extended. When $\lambda>2$, it monotonously
increases as the increasing of $p$. Those can be understood from
the varying of the average von Neumann entropy $\langle E_v^\alpha
\rangle$ of the individual eigenstate with $\lambda$ and the
number of shortcut links. Especially, at $\lambda>2$ we find there
exists the localization-delocalization transition of electron
states reflected from von Neumann entropy. In a word, the varying
of $\langle E_v \rangle$ with $p$ are similar to that for periodic
QSWNs at $\lambda<2$ and similar to that for disorder QSWNs at
$\lambda>2$.


\begin{acknowledgments}
This work is partly supported by the National Nature Science
Foundation of China under Grant Nos. 90203009 and 10175035, by the
Excellent Young Teacher Program of MOE, P.R. China, and by  of the
Foundation of Nanjing University of posts and telecommunications
under Grant No.NY205050, P.R. China.
\end{acknowledgments}

\end{document}